# Some proves of integrated influence of geomagnetic activity and weather changes on human health

*Khabarova O.V.[1], Dimitrova S.[2]*

[1] Space Research Institute (IKI) RAS, Moscow, Russia, olik3110@aol.com
[2] Solar-Terrestrial Influences Laboratory, Bulgarian Academy of sciences, Sofia, Bulgaria

Our environment includes many factors, and each person on the Earth is permanently influenced by two of them: weather and magnetic field. It was found in the works of many investigators that the weather changes correlate with human health state. In the same time, disturbances of geomagnetic field (as one of the space weather manifestations) may influence bioobjects, including people.

In this work we demonstrate the cumulative effect of different external factors (space weather and meteorological weather parameters) on human health on the base of medical experimental data (blood pressure and heart rate data rows for 86 people). It is shown that inclusion both solar-geomagnetic and weather parameters in simulation process give adjusting mixed parameter, which correlates with health state significantly better, than separated environmental parameters do.

## Introduction

Meteorological weather and space weather are two interrelated factors, permanently influencing human being [1, 2].

Sharp changes of meteorological parameters lead to increased morbidity level over the world [3, 4].

Space weather can cause changes in the Earth's magnetic field. Currently, an overwhelming majority of researchers, who investigate solar-biospheric relationships, agree with the electromagnetic nature of the effect of the Sun on biosphere. Magnetic field penetrates in human body freely, and field changes can change some its characteristics, because the organism is conductive medium and all processes in our organism are based on electricity.

It is proved at very good correlative level since A.L.Tchizjevsky that global and long-period (up to several years) changes in space weather and geomagnetic field cause global changes in biosphere (like increase/decrease of number of wars, conflicts, revolutions, human morbidity, intellectual and physiological activity of people and so on), see, for example, [5-7]. The correlation level between space weather/geomagnetic parameters and long-term medical or social statistics data is about 0.5÷0.8.

Meanwhile, short term space weather influence on people is not so obvious, and linear correlation level between bio-medical parameters and indices of geomagnetic activity is usually no more than ~ 0.3÷0.4 [8-9].

The problem particularly is related to the complexity of studied objects, impossibility to separate out the dominant factor from a number of others, absence of repeatability, insufficiency of statistics (data) and many other difficulties of purely technical character. However the biggest difficulties and observed artefacts are connected with the frequent use of the statistical analysis which is inapplicable for systems with floating (varying) time of the response (it is typical for all biological systems). Time of the response of bio-system to external influence can be various and the response itself - nonlinear.

Presence of reaction directly over the days of space (or meteorological) weather sharp changes, geomagnetic storms and other environmental changes is usually demonstrated by methods of fragmentation of solar and geomagnetic data according to different intensities, by superimposed epoch method, or method of hypothesis checking (when two rows are converted into rows of events "one-zero" and then checking of randomness of their coocurrence is evaluated) [8, 10-11].

Short-period variations of space weather and geomagnetic field are associated mainly with changes of the normal functioning of biological objects and produce a short-term increase in the number of cardiac and infectious diseases, traffic and industrial accidents, heart attacks and sudden cardiac deaths, cellular changes, etc. [12-14].

It is known that combination of different negative factors may lead to failure of adaptation and, as a consequence, to heritable and chronic diseases, especially for children [15]. This is so-called cumulative effect of negative influences.

Here we try to show that taking into account cumulative effect may be used for successful calculation (and possible prognosis) of human organism's reaction to external effects.

## Methods, data used, and previous results

The consequent daily measurements of heart rate, blood pressure and complaints number were performed by Dimitrova during the autumn 2001 and spring 2002 (92 days on the whole) for 86 volunteers (33 males and 56 females with an averaged age 48±12 years) in Sofia city (42 41N; 23 19E) as a part of experiment on space weather possible influence on human health.

Systolic blood pressure represents the maximum force exerted by the heart against the blood vessels during the heart's pumping phase. Diastolic pressure is the resting pressure during the heart's relaxation phase. Chronic hypertension is defined as a systolic blood pressure of 140 millimetres of mercury (mm Hg) or higher and a diastolic blood pressure of 90 mm Hg or higher. Meanwhile, blood pressure may vary from day to day for practically health persons as a result of human organism adaptation to some external and internal changes.

Previous results have shown that arterial blood pressure and subjective psycho-physiological complaints may increase with the increase of geomagnetic activity on the days prior and after geomagnetic storms main phase [8, 16, 17].

The average increment of systolic and diastolic blood pressure of the group examined reached 9%. This effect was present irrespectively of gender. Results obtained suppose that hypertensive persons have the highest sensitivity and the hypotensive persons have the lowest sensitivity of the arterial blood pressure to increase of geomagnetic activity. The results did not show significant changes in the heart rate. The percentage of the persons who reported subjective psycho-physiological complaints was also found to increase significantly with the geomagnetic activity increase and the highest sensitivity was revealed for the hypertensive females.

It was found that diastolic pressure is most sensitive parameter of cardio-vascular system to changes of external influences [18]. Reaction of human organism on weather changes is nonlinear and detected with difficulty, because of number of meteorological parameters, included in the "weather" concept. In [18] the weather strength parameter $S$, including six main meteorological parameters, was introduced into practice:

$$S = \frac{(2+100 \cdot Temperature) \cdot (1+10 \cdot Wind\ speed) \cdot (10 \cdot Cloudiness + Humidity)}{\Pr essure^2 \cdot Visibility}$$

$S$ allows to linearize a task and to reach the visible relations between bio-medical data, describing human health state, and weather parameter changes. Correlation between blood pressure and Weather strength parameter is negative (see Figure 1).

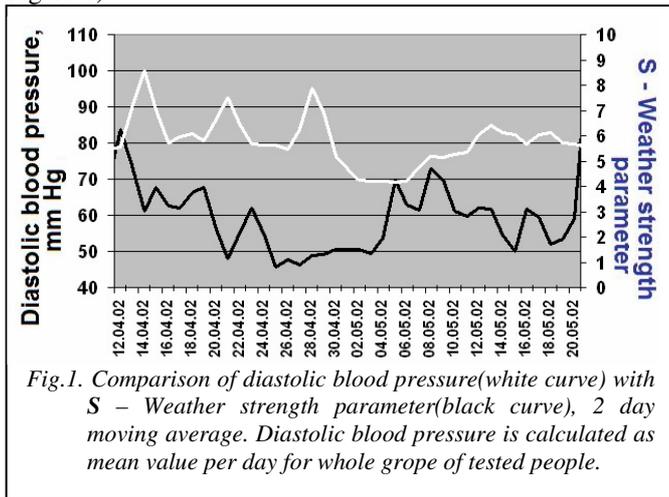

*Fig.1. Comparison of diastolic blood pressure(white curve) with S – Weather strength parameter(black curve), 2 day moving average. Diastolic blood pressure is calculated as mean value per day for whole grope of tested people.*

One of the approach to the study of the problem of space weather possible influence on human health is comparison of health parameters not only with standard indices of geomagnetic and solar activity (most of them have time resolution ≥1 hour), but also with geomagnetic pulsations presence or pulsation indices. Table of pulsation types and their possible connection with health can be found at: http://hypertextbook.com/facts/2001/ElizabethWong.shtml as a part of electronic "Physics Factbook".

It is possible to estimate the power of the geomagnetic fluctuations in ULF-range (2-10 mHz) with help of recent global ULF-index of geomagnetic activity. Kp, Dst, AE, SYMH, PC, IMF parameters, etc. quantify the laminar energy supply in the solar wind-magnetosphere-ionosphere system. ULF wave index characterizes the turbulent character of the energy transfer from the solar wind into the magnetosphere and the short-scale variability of near-Earth electromagnetic processes.

The ground ULF wave index is a proxy of global ULF activity is constructed using 1-min data from all available magnetic stations in Northern hemisphere: hourly band-integrated (2-10mHz) spectral power of two horizontal components. Database of hourly ULF indices is available on an anonymous FTP site ftp://space.augsburg.edu/maccs/ULF_index. See also information about this index in [19, 20].

Ptitsyna and others shown in 1998 that ultra low-frequency (ULF) low-intensity environmental magnetic fields affect the human nervous system, heart attacks are also may be associated with influence of magnetic field with ULF frequencies [21].

Otsuka in 2001 considered wide period branch of pulsations and concluded that link of low-frequency pulsations with heart variability is observed only in the season in which sunshine alternated with darkness, a finding suggesting a mechanism influenced by the alternation of light and darkness [22].

Khabarova in 2002 collected some previous results in this area and has shown that they can be explained by adaptation mechanism's starting due to parametrical resonant influence on human brain [23].

Recent investigations [24] show that geomagnetic micropulsations Pc1 (0.2-5s) could be also bioeffective for cardiovascular system and modulate the number of myocardial infarctions and sudden death, especially during the winter (this result is well-correlated with result by Otsuka et al.).

Unfortunately, number of works in this area is very limited and additional experiments are obviously needed.

Here we have used both S-weather strength parameter and ULF-index database for comparison with cardiovascular parameters.

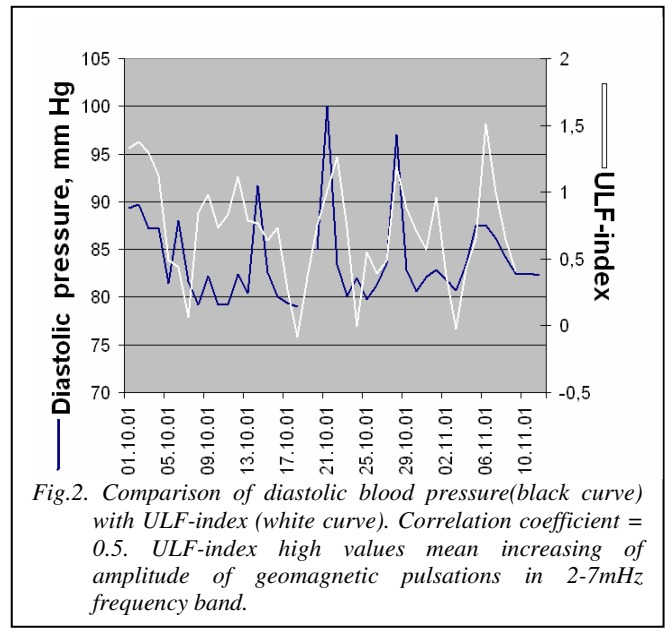

*Fig.2. Comparison of diastolic blood pressure(black curve) with ULF-index (white curve). Correlation coefficient = 0.5. ULF-index high values mean increasing of amplitude of geomagnetic pulsations in 2-7mHz frequency band.*

## Results

As we have mentioned, the linear correlation coefficients between various geomagnetic indices and medico-biological parameters rarely exceed 0.4. In this investigation it was unexpectedly found enough high level of correlation between ULF-index and cardiovascular parameters data (it equals 0.5 during the autumn and 0.4 during the spring period of measurements). Example can be viewed in Figure 2.

The question appears: Is it the maximum obtainable level of correlation or we can simulate cardiovascular parameters better?

We have tried to find some adjusting parameter, which include most number of reasonable external parameters.

In addition to ULF-index, the Kp-index of geomagnetic activity, Sunspot number, and *S* were examined. Simulation give the best results, when adjusting parameter WGS (weather, geomagnetic, solar) is as follow:

$$WGS = \frac{Kp + ULF \cdot Wolf/3}{7 + S}, \qquad (1)$$

where Kp is geomagnetic Kp index; ULF is geomagnetic ULF-index, *S* – weather strength parameters. All these parameters included in *WGS*, have approximately the same weight. Result is given in the Figure 3.

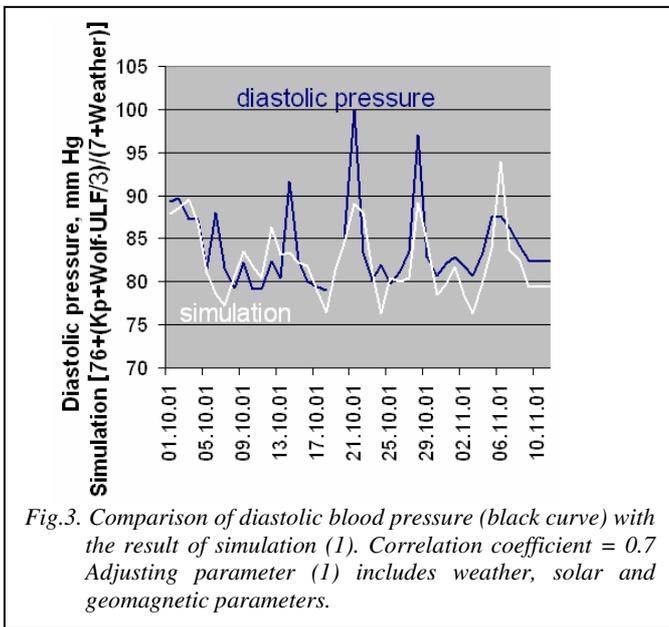

*Fig.3. Comparison of diastolic blood pressure (black curve) with the result of simulation (1). Correlation coefficient = 0.7 Adjusting parameter (1) includes weather, solar and geomagnetic parameters.*

Figure 4 provides information on correlation coefficient level (for all data base on blood pressure) between systolic and diastolic pressure and parameters, included in *WGS* adjusting mixed parameter.

It is easy to see that adjusting parameter *WGS* is the best correlated with blood pressure, although some parameters, included in it, have low (but statistically significant) correlation levels.

Correlation between *WGS* and diastolic blood pressure during the autumn period is 0.65, during the spring is 0.50, and it equals 0.54 for whole period of investigation.

## Conclusions and discussion

It was found out that including different parameters, possibly influenced people health, into consideration and simulation give the result better in comparison with attempts to separately correlate these potentially bioeffective factors with health parameters.

Usually the most reliable linear correlative tie between geomagnetic field intensity and human health state exists in high latitudes, but most people live in the latitudes, where this tie is just probabilistic [11, 18].

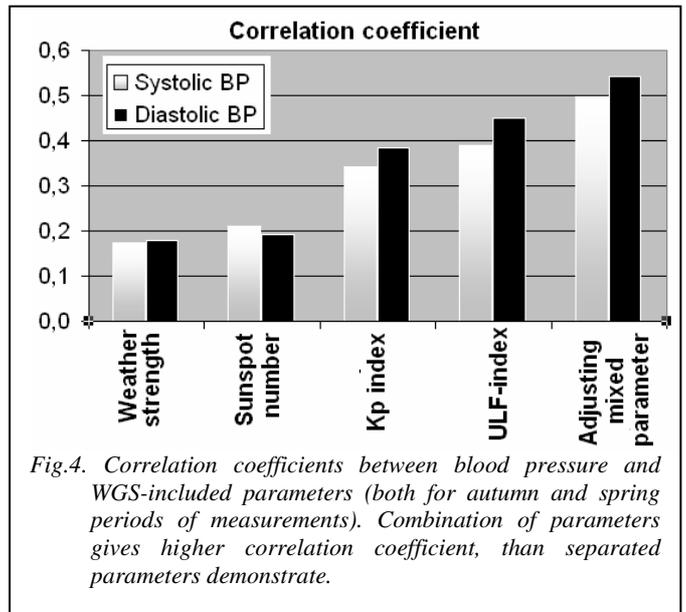

*Fig.4. Correlation coefficients between blood pressure and WGS-included parameters (both for autumn and spring periods of measurements). Combination of parameters gives higher correlation coefficient, than separated parameters demonstrate.*

We have considered here geomagnetic and solar indices, and weather strength parameter, including six main meteorological parameters. Their combination gives correlation level, exceeding the levels, typical for each parameter.

It is obvious that even high correlation coefficient and coincidence of events do not prove direct influence of one factor to other, but successful simulation may give some confidence in the right way of investigations and really affecting factor searching.

Cumulative effect of different negative external influences may became apparent in the absence of high correlations between separated parameters, but their integrated effect may be obvious and good-observed even at correlative level.

Many processes on the Earth and in space are marginally connected with each other and in the same time are different, but all they influence people every time, hence their possible bioeffecteveness must be investigated together.

We have shown here that increasing of the number of considered bioeffective parameters can lead to possibility of the adequate simulation of observed medico-biological data rows and, in the future, to prediction of negative biological effects of environmental changes.